\documentclass{iaus}
\usepackage{graphics}

  \checkfont{eurm10}
  \iffontfound
    \IfFileExists{upmath.sty}
      {\typeout{^^JFound AMS Euler Roman fonts on the system,
                   using the 'upmath' package.^^J}%
       \usepackage{upmath}}
      {\typeout{^^JFound AMS Euler Roman fonts on the system, but you
                   dont seem to have the}%
       \typeout{'upmath' package installed. iaus.cls can take advantage
                 of these fonts,^^Jif you use 'upmath' package.^^J}%
      }
  \else
  \fi

  \checkfont{msam10}
  \iffontfound
    \IfFileExists{amssymb.sty}
      {\typeout{^^JFound AMS Symbol fonts on the system, using the
                'amssymb' package.^^J}%
       \usepackage{amssymb}%

      }{}
  \fi

  \IfFileExists{amsbsy.sty}
    {\typeout{^^JFound the 'amsbsy' package on the system, using it.^^J}%
     \usepackage{amsbsy}}
    {\providecommand\boldsymbol[1]{\mbox{\boldmath $##1$}}}

\newsavebox{\astrutbox}
\sbox{\astrutbox}{\rule[-5pt]{0pt}{20pt}}

\newcommand{\Lo}{\mbox{L$_{\odot}$}}
\newcommand{\kms}{\mbox{km~s$^{-1}$}}
\newcommand{\x}{\mbox{$\times$}}
\newcommand{\uSFR}{\mbox{M$_{\odot}$~yr$^{-1}$}}

\title[]
      {Luminous Infrared Starbursts in a Cluster of Galaxies}

\author[]
{Pierre-Alain Duc$^{1}$, Dario Fadda$^{2}$, Bianca Poggianti$^{3}$, \break David Elbaz$^{1}$, Andrea Biviano$^{4}$, Hector Flores$^{5}$ \break  Alan Moorwood$^{6}$, Alberto Franceschini$^{7}$ \and Catherine Cesarsky$^{6}$}

\affiliation{$^{1}$CEA--Saclay and CNRS FRE 2591, France, email:paduc@cea.fr\\
$^{2}$Caltech, SIRTF Science Centre, Pasadena, USA\\
$^{3}$Osservatorio Astronomico di Padova,  Padova, Italy\\
$^{4}$INAF/Osservatorio Astronomico di Trieste, Trieste, Italy\\
$^{5}$Observatoire de Paris, GEPI,  Meudon, France\\
$^{6}$European Southern Observatory,  Garching, Germany\\
$^{7}$Dipartimento di Astronomia,  Padova,  Italy}

\pubyear{2004}
\volume{195}
\pagerange{1--8}
\date{?? and in revised form ??}
\jname{Outskirts of Galaxy Clusters: intense life in the suburbs}
\editors{A. Diaferio, ed.}
\begin{document}

\maketitle

\begin{abstract}
Analysing mid--infrared ISOCAM  images of the cluster of galaxies  J1888.16CL, we  identified among
its members several particularly active 
galaxies  with total infrared luminosities well above  $10^{11}~\Lo$.
If powered by dust enshrouded starbursts, as suggested by their optical spectra, 
these Luminous Infrared Galaxies would exhibit  star  formation rates surprisingly high in the cluster
environment. The triggering mechanism  is unclear but could be tidal collisions
within sub-structures or infalling groups.
\end{abstract}


\section{Introduction}
In the Universe, the galaxies  forming stars with 
the highest rates and/or exhibiting the most extreme nuclear radiation turn out 
to be the most obscured ones. Their  internal activity  being partly  dust-enshrouded, it remains underestimated 
in visible observing bands and  shows up in the infrared regime.  Such so-called Luminous Infrared 
Galaxies (LIRGs) seem  to be more prevalent  in relatively dense environments
 like groups where tidal galaxy interactions  play a role in enhancing their activity.
On the other hand, the densest regions like clusters are
  usually considered as being particularly unfavorable for the global activity
of their host galaxies. The combined action of  multiple high-speed collisions, of the global cluster tidal field, and of the 
ram-pressure exerted by  the rich intra-cluster medium  contribute  to strip galaxies  of their disk and halo 
gas reservoirs. Without this fuel, star formation is  quenched 
and eventually stops while  the galaxy center is no longer fed and  any  nuclear 
activity ends. 

\vspace{0.3cm}

We present here   data obtained with the ISOCAM camera on board the Infrared Space 
Observatory (ISO) that reveals  the presence 
of confirmed cluster members with  high infrared luminosities and inferred star formation
 rates of  several tens of solar masses per year.  These LIRGs were
discovered in the X-ray luminous cluster J1888.16CL, situated at a redshift of 0.56. The observations 
performed in the LW3 band centered at 15 $\mu$m
covered a strip of $3 \times 15$ square arcminutes.
Follow-up ground based observations were carried out using ESO
facilities in Chile. They consist of optical B,V,R WFI/2.2m  images, 
near-infrared J,H,Ks SOFI/NTT  images and  optical FORS1/VLT multi-slit
spectra. This collection of data  allowed us to assess the cluster membership of the
ISO sources and study their spectral properties.

\begin{figure}
\centerline{\resizebox{\textwidth}{!}{\includegraphics{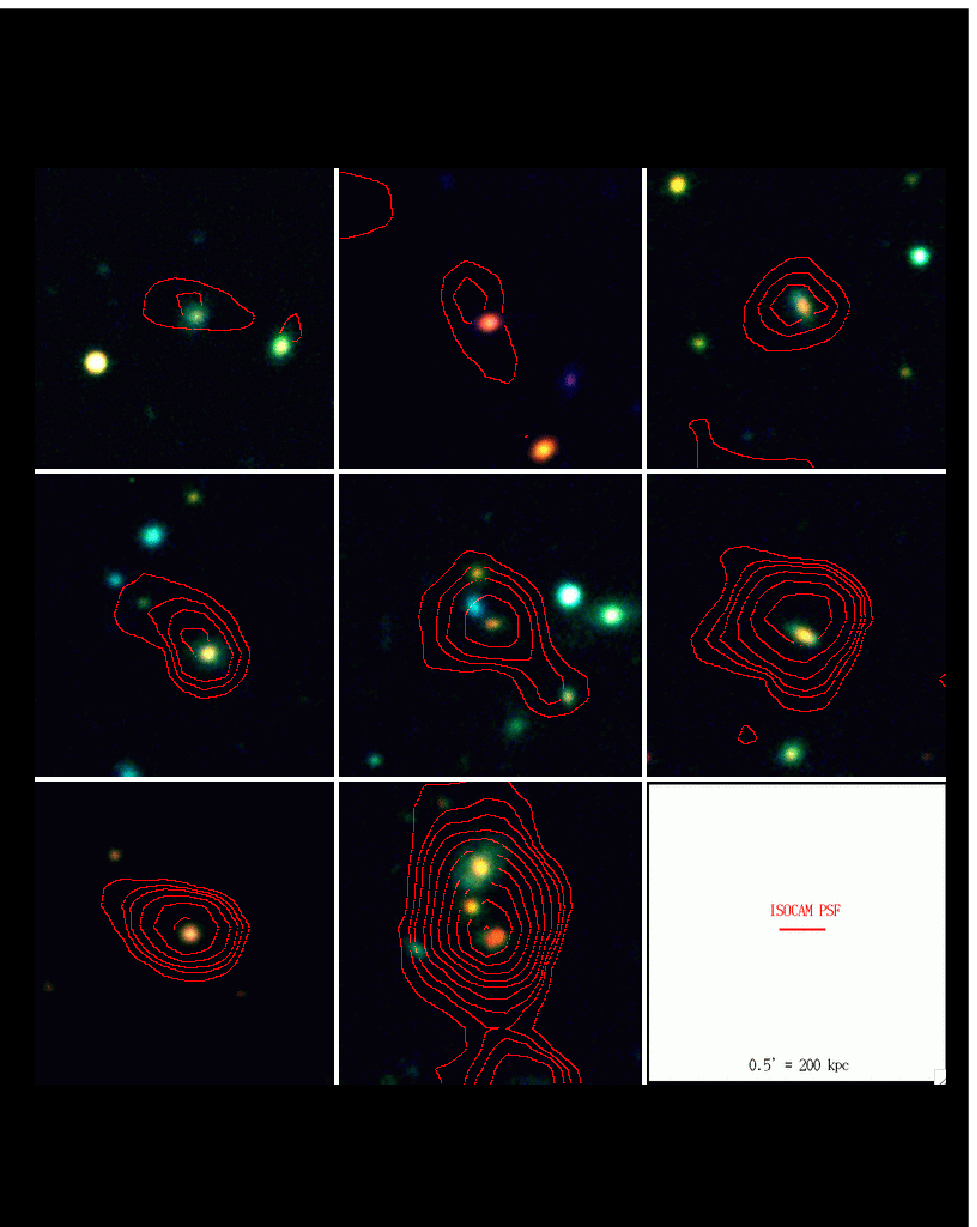}}}
  \caption{Mid-infrared emitters towards the rich 
cluster of galaxies  J1888.16CL. The 15 $\mu m$  emission is 
shown with   contours corresponding to different signal to noise ratio levels
starting with a SNR of 3.      
In this figure are  presented the six 
ISOCAM emitters firmly identified with  galaxies having a measured spectroscopic
redshift consistent with  cluster membership. The two bright sources at the
bottom may lie  in groups falling towards the cluster.
}
\end{figure}

\section{LIRGs in the cluster J1888.16CL}

A total of 44 sources were detected at 15 $\mu$m and  the vast majority were
unambiguously identified on optical/near-infrared images, thanks to the
fair spatial resolution of ISOCAM (4.6" at 15 $\mu$m;  see Fig.~1). Among the 27 objects
for which spectra could be obtained, six have a redshift consistent with 
cluster membership while two other sources with slightly higher redshifts may lie in infalling groups.
The photometric 
redshift of all remaining mid-infrared emitters are consistent with a foreground or 
background location. The most distant source at a redshift of 1.7 is an atypical   hyperluminous
broad absorption line quasar (\cite[Duc et al., 2002]{Duc02}).
The ISO sources in the cluster  are uniformly distributed along the 6 \x 1.3 Mpc$^2$ strip
mapped by ISOCAM and have a velocity dispersion of $700  ^{+520}_{-260}~\kms$,
 which is marginally comparable to that measured  for all cluster members ($450  ^{+110}_{-80}~\kms$).

The confirmed cluster members detected by ISOCAM have  15  $\mu$m  fluxes ranging between 
200 and 750  $\mu$Jy. We estimated their total infrared luminosity using the mid-infrared
to far-infrared correlation that has been well established for nearby  galaxies observed
with the IRAS satellite  (\cite[Chary \& Elbaz 2001]{Chary01}). All sources  have an inferred
IR luminosity above $1.3 \times 10^{11} ~\Lo $, and hence may be considered
as genuine Luminous Infrared Galaxies (see Fig.~2). 
Either  nuclear activity or star formation could  heat the dust responsible for
the infrared emission. Recently, a  high number of  galaxies
shining in the hard X-rays (\cite[Martini et al., 2002]{Martini02}; \cite[Johnson et al., 2003]{Johnson03}) and in the radio centimetric regime
(\cite[Morrison \& Owen 2003]{Morrison03}; \cite[Best, 2004]{Best04})  was found  in  Chandra images and  Very Large Array maps
of distant clusters, respectively. They are  suspected to host an AGN.
However, studies  based on infrared data obtained with  ISO  and
X-ray data collected by  XMM-Newton and Chandra  indicate that, statistically,  
infrared-selected luminous galaxies   are  predominantly powered by star formation, at least in 
the field environment  (\cite[Fadda et al., 2002]{Fadda02}).
All J1888 ISO sources have weak  emission lines in their optical spectrum; none of the
latter are   broad enough to be classified as type 1 AGNs. They typically exhibit
an [OII]3727 emission line with a moderate equivalent width 
(less than 40\AA) and strong early Balmer Hydrogen absorption lines (the equivalent width of H$\delta$ is above 4\AA).
This  spectroscopic signature 
is best explained by spectrophotometric models including a selective 
dust extinction depending on the stellar age and a high rate of current star formation
(\cite[Poggianti et al., 2001]{Poggianti01}).  
Such an optical signature gives further support to the hypothesis that 
the infrared emission detected in the cluster results from 
 a  dust enshrouded starburst. From the inferred total infrared 
luminosities, one may deduce  individual star formation rates (SFR) ranging between 20 and 120 \uSFR~(a maximum of 270 \uSFR\ including the two ULIRGs for which the cluster membership is not  assessed)

 \begin{figure}
\centerline{\resizebox{\textwidth}{!}{\rotatebox{270}{\includegraphics{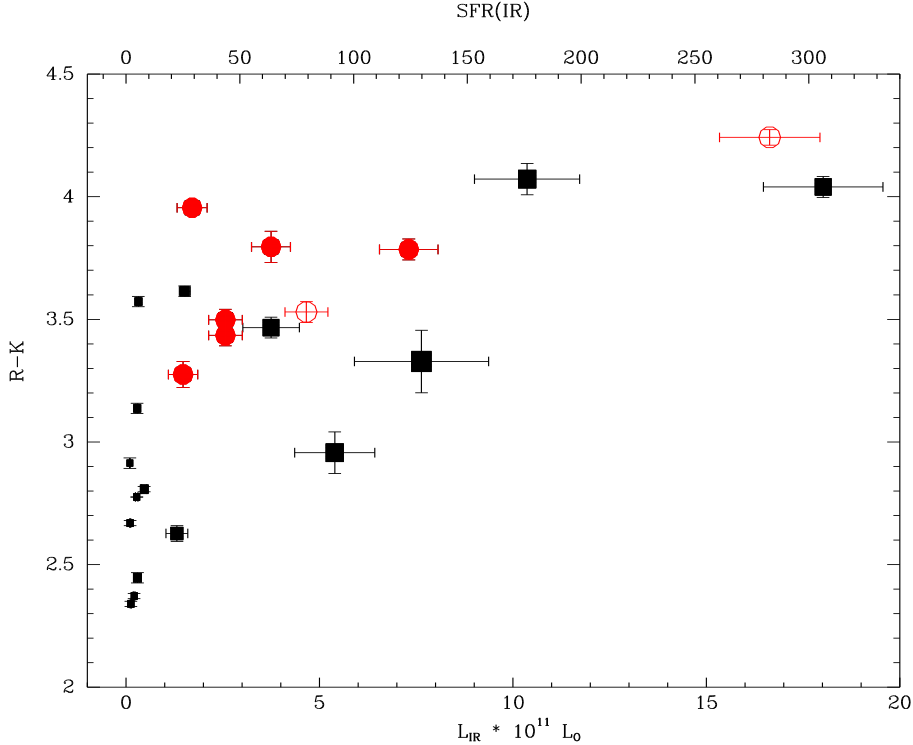}}}}
  \caption{Photometric R-K color index versus the inferred Infrared luminosity
 of the ISOCAM-detected galaxies towards the field of J1888.16CL. 
The size of each dot is proportional to the galaxy redshift.
The  Star Formation Rates in solar masses  per year are indicated along the top
axis. They were computed from the mid--infrared luminosities.
Confirmed cluster members are shown with (red) dots while the two galaxies likely
associated with infalling groups  are indicated with the two empty circles. The black
filled squared are either foreground or background galaxies.
}
\end{figure}

\section{Triggering the LIRG/starburst phase in the cluster}

The discovery of extremely active galaxies in  X--ray luminous clusters  was 
a priori unexpected given all the processes that eventually inhibit star formation in such
an environment. 
Optical  surveys of local clusters indicate a lack of 
galaxies with blue colors or exhibiting emission lines in their  spectrum and hence 
showing signs of  star formation activity. Such galaxies are however more numerous 
in more distant and younger clusters (\cite[Dressler et al., 1999]{Dressler99}). Another evolutionary effect  is the 
increased number of galaxies with spectra characterized 
by strong Balmer absorption lines and no emission lines. This signature is consistent
with that expected for galaxies having suffered a recent but now faded
intense star formation episode (\cite[Poggianti et al., 1999]{Poggianti99}). This interpretation however
was challenged (\cite[Balogh et al., 1999]{Balogh99}) since prompt truncated star formation 
may have the same effect in some cases. The presence of  post-starburst
galaxies implies that starbursts may occur in clusters; they would have remained unnoticed 
in optical surveys because of their dust cocoon.  

What triggered the enhanced infrared activity in the ISO J1888 galaxies ?
The answer will very much depend on when and where the LIRG event
was switched on.

One simple hypothesis is that the mid-infrared emitters have recently
been accreted by the cluster and that the luminous infrared phase actually started in the
coeval field where LIRGs  are not uncommon. Under this assumption, the infall rate 
that would account for the number of detected galaxies is constrained by the duration of 
the infrared luminous phase which cannot exceed 100 Myr. During that period at least six field 
LIRGs should have been accreted. 
The proportion  of LIRGs in the field is still poorly known\footnote{whether  the field actually
exists as such is even questionable} but could be of the order of 5--10 \% in the redshift interval 0.5-0.6 
(\cite[Flores et al., 1999]{Flores99b}). The total inferred  infall rate (including the IR-quiet galaxies) would then be
of about 100 massive galaxies in 100 Myr, which seems unrealistic.
Numerical simulations and X--ray observations show however that accretion onto clusters 
from the field is not a  spherically  symmetric process, but occurs along filaments  or via
 mergers with other groups and  clusters.  One therefore cannot exclude the possibility that the LIRGs 
observed  in J1888.16CL  belonged to such a recently accreted structure.
where tidal interactions could be an efficient mechanism in triggering the star formation.

The alternative is that the luminous infrared phase and associated starburst
event occurred within the cluster. There the large velocity dispersion does not
favor  tidal galaxy-galaxy collisions and mergers. 
 Ram-pressure may create some shocks and instabilities in the gaseous component while it is stripped; 
the changing global cluster tidal field may destabilize the galaxy disks,  when 
they lie in infalling groups (\cite[Bekki, 1999]{Bekki99}) or when a cluster-cluster merger is involved
(\cite[Miller \& Owen, 2003]{Miller03}). Such mechanisms can account for  
 a mild star formation episode but are probably not efficient enough   to concentrate 
large quantities of gas in the nuclear regions, as required to trigger a starburst or
  an AGN activity with an intensity comparable to the local LIRGs.

\section{Conclusions}

Although the effective triggering mechanism remains unclear  --perhaps tidal interactions
in sub-structures --, our observations clearly indicate
that  a significant activity may occur in  clusters, even at
moderate redshifts and hence a long time after their collapse and the formation
of the bulk of their stellar populations. This result is corroborated with other studies
based on ISOCAM mid--infrared data  (Ricardo-Perez et al., in these proceedings), optical data
(see the review by A. Dressler), radio continuum maps and perhaps millimetric SCUBA 
surveys   (Webb et al., these proceedings; \cite[Best, 2002]{Best02}).
The MIPS camera on board  the infrared satellite Sptizer will also very soon provide ample 
information on the dust  enshrouded activity in clusters.
 This secondary peak of activity may correspond to a phase in the evolution of
clusters when accretion of galaxies, especially through the merging of groups, was 
 conspicuous.

\end{document}